# A Smartphone enabled Approach to Manage COVID-19 Lockdown and Economic Crisis

Halgurd S. Maghdid, *Kayhan Zrar Ghafoor

*Abstract*—The emergence of novel COVID-19 causing an overload in health system and high mortality rate. The key priority is to contain the epidemic and prevent the infection rate. In this context, many countries are now in some degree of lockdown to ensure extreme social distancing of entire population and hence slowing down the epidemic spread. Further, authorities use case quarantine strategy and manual second/third contact-tracing to contain the COVID-19 disease. However, manual contact tracing is time consuming and labor-intensive task which tremendously overload public health systems. In this paper, we developed a smartphone-based approach to automatically and widely trace the contacts for confirmed COVID-19 cases. Particularly, contact-tracing approach creates a list of individuals in the vicinity and notifying contacts or officials of confirmed COVID-19 cases. This approach is not only providing awareness to individuals they are in the proximity to the infected area, but also tracks the incidental contacts that the COVID-19 carrier might not recall. Thereafter, we developed a dashboard to provide a plan for government officials on how lockdown/mass quarantine can be safely lifted, and hence tackling the economic crisis. The dashboard used to predict the level of lockdown area based on collected positions and distance measurements of the registered users in the vicinity. The prediction model uses K-means algorithm as an unsupervised machine learning technique for lockdown management.

*Index Terms*—COVID-19, contact-tracing, GPS positioning, Smartphone.

## I. INTRODUCTION

In an unprecedented move, China locks down the megacity named Wuhan, in which the novel coronavirus was first reported, in the hopes stopping the spread of deadly coronavirus. During the lockdown, all railway, port and road transportation were suspended in Wuhan city. With the increasing number of infections and fast person-to-person spreading, hospitals are overwhelmed with patients. Later, the disease has been identified in many other countries around the globe [1], [2]. Subsequently, the World Health Organization (WHO) announced that the virus can cause a respiratory disease with clinical presentation of cough, fever and lung inflammation. As more countries are experienced dozens of cases or community transmission, WHO characterized COVID-19 disease as a pandemic.

Halgurd S. Maghdid is with the Department of Software Engineering, Faculty of Engineering, Koya University, Kurdistan Region-F.R.Iraq. First.Last@koyauniversity.org.

Kayhan Zrar Ghafoor is with the Department of Software Engineering, Salahaddin University-Erbil, Iraq; School of Mathematics and Computer Science, University of Wolverhampton, Wulfruna Street, Wolverhampton, WV1 1LY, UK. kayhan@ieee.org.

*Kayhan Zrar Ghafoor is the corresponding author. kayhan@ieee.org.

The researchers can access the implementation and programming code in https://github.com/halgurd18/lockdown_COVID19

In such unprecedented situation, doctors and health care workers are putting their life at risk to contain the disease. Further, in order to isolate infected people and combatting the outbreak, many hospitals are converted to COVID-19 quarantine ward. Moreover, a surge of COVID-19 patients has introduced long queues at hospitals for isolation and treatment. With such high number of infections, emergency responders have been working non-stop sending patients to the hospital and overcrowded hospitals refused to in more patients. For instance, recently in Italy medical resources are in short supply, hospitals have had to give priority to people with a significant fever and shortness of breath over others with less severe symptoms [3].

As the COVID-19 continues to spread, countries around the glob are implementing strict measures intensify the lockdown, from mass quarantine to city shutdown, to slow down the fast transmission of coronavirus [4]. During the lockdown, people are only allowed to go out for essential work such as purchasing food or medicine. Ceremonies and gatherings of more than two people are not permitted. These strict rules of quarantine that only allows few to move around the city including delivery drivers providing vital lifeline. On the other hand, few countries, such as Japan, has declared a state of emergency in many cities in an attempt to tackle the spread of the virus. Although COVID-19 started as a health crisis, it possibly acts as a gravest threat to the world economy since 2008 global financial crisis [5].

COVID-19 epidemic affect all sectors of the economy from manufacturing and supply chains to universities. It is also affect businesses and daily lives especially in countries where the COVID-19 has hit the hardest. The shortage of supply chain has knock-on effects on economic sector and the demand side (such as trade and tourism). This makes a supply constraint of the producer and causing a restraint in consumer's demand, this may lead to demand shock due to psychological contagion. In order to prevent such widespread fallout, central banks and government have been rolling out emergency measures to reassure businesses and stabilize financial markets to support economy in the phase of COVID-19. Currently, most countries are in the same boat with leading responsibility of Group Twenty and international organizations [5]. To meet the responsibility, many companies and academic institutions around the world made efforts to produce COVID-19 vaccine. But, health experts stating that it may take time to produce an effective vaccine.

As an effective vaccine for COVID-19 isn't probably to be in market until the beginning of next year, management of lockdown is an imperative need. Thus, public health officials combat the virus by manual tracking of recent contacts history of positive COVID-19 cases. This manual contact tracing is very useful at the early spreading stage of the virus. However, when the number of confirmed cases was increased tremendously in some countries, manual contact tracing of each individual is labor-intensive and requires huge resources [6]. For example, an outbreak of the COVID-19 at a funeral ceremony in an avenue in Erbil, Kurdistan Region left regional government with hundred of potential contacts. This situation or many other scenarios of massive number of cases burden the government on trying to manual tracking all contacts [7]. It is risky that health authorities cannot easily trace recent COVID-19 carrier cases so that its probability of occurrence and its impact can hardly be measured.

Technology can potentially be useful for digital contact-tracing of positive coronavirus cases. Smartphone can use wireless technology data to track people when they near each other. In particular, when someone is confirmed with positive COVID-19, the status of the smartphone will be updated and, then the app will notify all phones in the vicinity. For example, if someone tests positive of COVID-19 and stood near a person in the mall earlier that week. The COVID-19 carrier would not be able to memorize the person's name for manual contact tracing. In this scenario, the smartphone contact-tracing app is very promising to notify that person [8]. This automated virus tracking approach could really transform the ability governments and health authorities to contain the and control the epidemic. In this situation, a dashboard is required to assist governments and health authorities to predict when lockdown and self-quarantine will end.

This research first reviews the state-of-the-art solutions to combat COVID-19. Then, we developed a smartphone-based approach to automatically and widely trace the contacts for confirmed COVID-19 cases. Particularly, contact-tracing approach creates a list of individuals in the vicinity and notifying contacts or officials of confirmed COVID-19 cases. This approach is not only providing awareness to individuals they are in the proximity to the infected area, but also tracks the incidental contacts that the COVID-19 carrier might not recall. Thereafter, we developed a dashboard to provide a plan for government officials on how lockdown/mass quarantine can be safely lifted, and hence tackling the economic crisis.

From a technical standpoint, we summarise the most important contributions of this paper as follows:

1) We build a tracking model based on positional information of registered users to conduct contact-tracing of confirmed COVID-19 cases.
2) We propose a smart lockdown management to predict a duration of lockdown.
3) In order to notify contacts for confirmed cases, we also developed a notification model to cluster lockdown regions.

The rest of this paper is organized as follows. Section II provide the literature review on recent advances of developed AI systems for COVID-19 detection. This is followed by presenting an overview of the proposed approach and details of the designed algorithm in section III. Section IV presents the experiments which are conducted in the paper. Finally, Section IV concludes the paper.

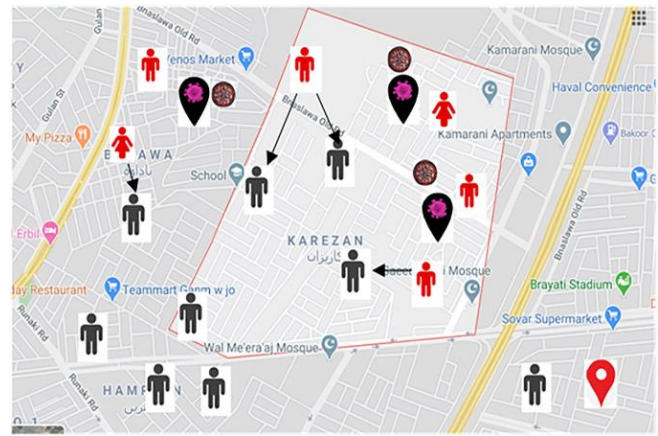

Fig. 1. Identifying and alerting people (black color) who have come into contact with a person (red color) infected with COVID-19

## II. BACKGROUND

In [9] the authors modeled on how COVID-19 spreads over populations in countries in terms of the transmission speed and containing its spreading. In the model, R is representing the reproduction number, which is defined the ability of the virus in infecting other people as a chain of contagious infection. Infected individuals rapidly infect a group of people over very short period of time, which then yields an outbreak. On the contrary, the infection would be in control if the probability gets closer of one person to infect less than one other person. This is exactly happening in Fig. 1; when people (black color) who have come into contact with an infected person (red color), the infection would be spread rapidly.

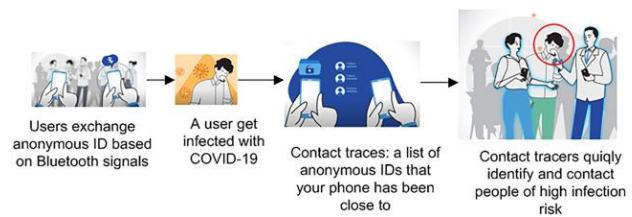

Fig. 2. Community-driven contact-tracing based on Bluetooth signal

One important aspect is how the number of infected people looks like depends on several factors, such as the number of vulnerable people in the communities, The time takes to recover a person without symptoms, the social contacts and possibility of infecting them with coronavirus. Further, another factor will affect fast spreading of coronavirus is the frequency of visiting crowded places such as malls and mini-markets. Thus, governments and public health authorities are responsible to manage and plan a convenient way to contain the epidemic. Moreover, countries at the early stage of virus spreading need to control the epidemic by typically isolating and testing suspected cases tracing their contact and quarantine those people in case they are infected. Testing and contact tracing at wide scale, the better the chance of containment.

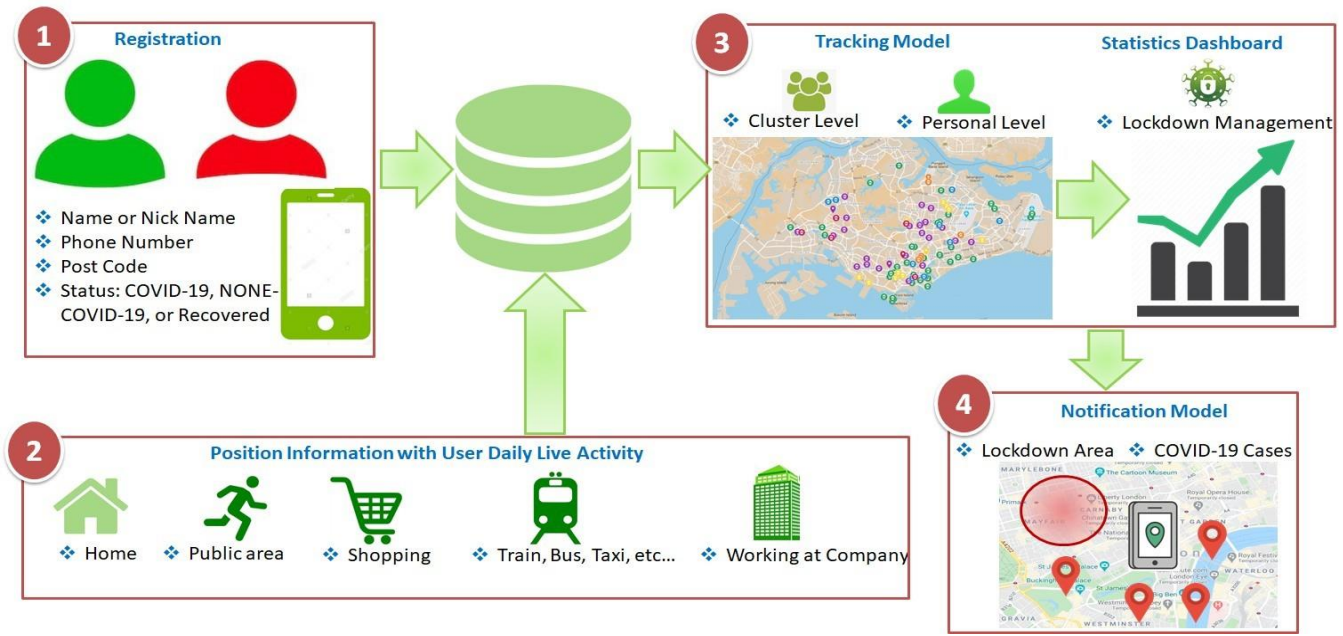

Fig. 3. A framework of contact-tracing using smartphone-based approach

In the case of COVID-19, research studies have been conducted for containment or controlling the fast spreading, and hence helping governments and societies in ending this epidemic. In [10], the authors have investigated the importance of confirmed COVID-19 case isolation that could play a key role in controlling the disease. They have utilized a mathematical model to measure the effectiveness of this strategy in controlling the transmission speed of COVID-19. To achieve this goal, a stochastic transmission model is developed to overcome the fast person-to-person transmission of COVID-19. According to their research study, controlling virus transmission is within 12 weeks or by a threshold of accumulative 5000 cases. However, controlling the spread of the virus using this mathematical approach is highly correlated to other factors like pathogen and the reaction of people.

One key role to track infected people and predict ending lockdown is contact-tracing. When a patient is diagnosed with infectious disease like COVID-19, contact-tracing is an important step to slowing down the transmission [11]. This technique seeks to identify people who have had close contact with infected individuals and who therefore may be infect themselves. This targeted strategy reduces the need for stay at home periods. However, manual contact tracing is subject to a person's ability to recall everyone they have come in contact over a two week's period. In [11], the authors exploited the cellphone's Bluetooth to constantly advertise the presence of people. These anonymous advertisements, named chirps in Bluetooth, are not containing positional or personally identifiable information. Every phone stores all the chirps that it has sent and overheard from nearby phones. Their system uses these lists to enable contact-tracing for people diagnosed with COVID-19. This system is not only traces infected individuals, but it also estimates distance between individuals and amount of time they spent in close proximity to each other. When a person is diagnosed with COVID-19, doctors would coordinate with the patient to upload all the chirps sent out by their phone to the public database. Meanwhile, people who have not been diagnosed can their phones do a daily scan of public database, to see if their phones have overheard any of the chirps used by people later diagnosed by COVID-19. This indicates that they were in close prolonged contact with that anonymous individual. Fig. 2 shows the procedure of exchanging anonymous ID among users for contact-tracing.

III. PROPOSED SMARTPHONE-BASED CONTACT-TRACING

As stated in the aforementioned section, manual contact-tracing is labor-intensive task. In this section, we detail out each part of the proposed smartphone-based digital contact-tracing shown in Fig. 3. The main idea of the proposed framework in Fig. 3 to enable digital contact-tracing to end lockdown and the same time preventing the virus from spreading. The best thing to do seems to be let people go out for their business, but any body tests positive of COVID-19, we would be able, through proposed framework, to trace everybody in contact with the confirmed case and managing the lockdown and mass quarantine. This will confirm preventing the spread of the virus to the rest of the people.

The first step of the proposed contact-tracing model is registration of users. There is no doubt registration and coverage of high percentage of population are very significant for effective pandemic control. Users provide information such as name, phone number, post code, status of the COVID-19 disease (Positive, Negative or recovered). Effectiveness of the application and digital contact tracing depends on two factors speed and coverage. For the proposed framework, we utilize Global Navigation Satellite System (GNSS) receiver for outdoor environment whereas Bluetooth low energy is used in indoors. Speed depends on how to reduce the time required for contact tracing from few days to hours or minutes. The more people register in the system, the better performance of the system in terms of both speed and coverage of contact tracing.

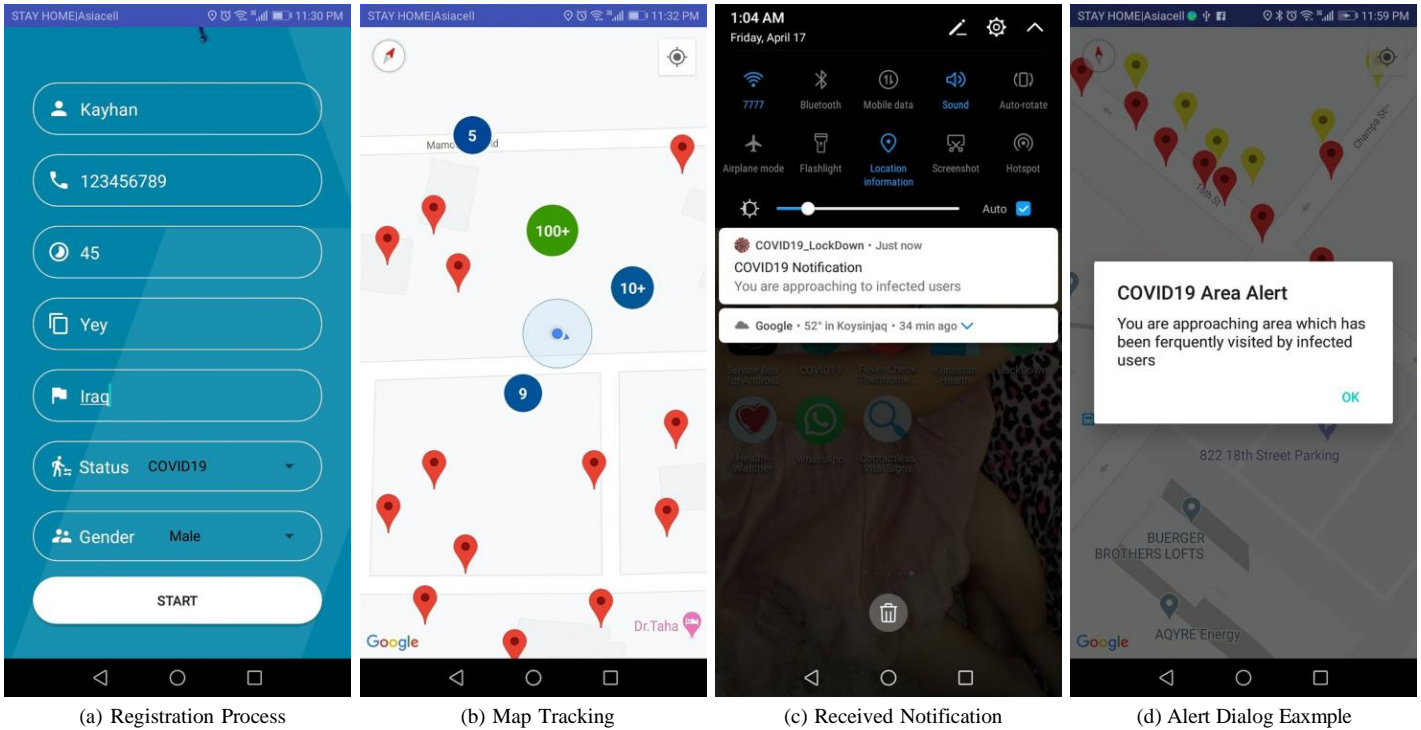

(a) Registration Process  (b) Map Tracking  (c) Received Notification  (d) Alert Dialog Eaxmple

Fig. 4. The consequent steps of the smartphone application.

In the second step, Global Positioning System (GPS) receiver is used by the proposed model to track either individuals or a group of people visiting to a common place. The GPS service class updates user coordinates to the database in every few seconds. Once a registered user reports gets infected with COVID-19, his test result would be send to the public database in central computer server. Other registered users will regularly check those central server provider for possible positive COVID-19 cases they were in contact in the past 2 weeks. Server is responsible to compare the infected ID with its list of stored IDs. A push notification will be send, by the server, to those who were in contact with a person tests positive. It is important to note that the information would be revealed to the central server is an ID of the phone.

Firebased cloud messaging is used to send push notification to multiple devices even the apps are paused or running in the background. Many apps send push notification, which indicate an alert to the users. This is happen when a person is approaching someone who is infected with COVID-19 or nearby a lockdown area. In order to protect the privacy of those who have the coronavirus, we only include an alerting message into the push notification. This certainly would be very useful for entire population to make informed decision about not getting close to COVID-19 area. However, this notification would help the public health professionals rather than replace it.

The proposal is also including a lockdown prediction model.

The model is working based on the collect geographic information and crowding level of the registered users in the system. In this study, K-means as an unsupervised machine learning algorithm is used to cluster the users' positions information and predict that the area should be locked down or not based on same empirical thresholds. Both scenarios results are shown in fig. 5.

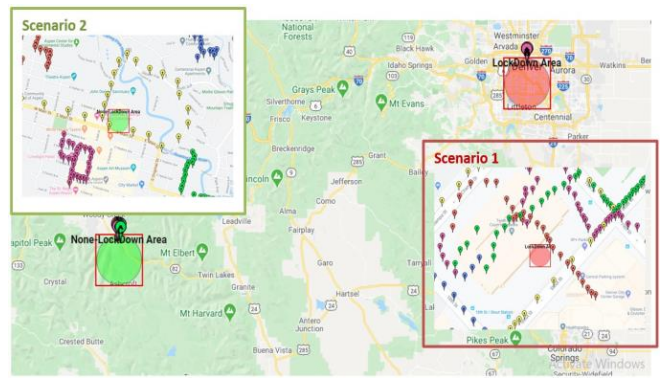

Fig. 5. The results of the prediction model for both scenarios.

## IV. EXPERIMENTS AND DEPLOYMENT

This section presents the details of how the proposed approach will be implemented. The proposal includes two main parts. First, deploying an application on Android-based smartphone which will be used by the users and track/send mobility information of the users to the system. While the second side is a web-portal (including a comprehensive dashboard) to monitor and predict the visited area that should be locked down or not.

### A. Smartphone Application

1- An Android application is implemented on the smartphone. The application lets the users to register their information into the proposed system including name, postcode or zip code, phone number, age, Bluetooth MAC address, gender, and COVID-19 status. The Bluetooth MAC address is automatically captured through the application without user interaction. The COVID-19 status includes three options which might be COVID-19, None COVID-19, and recovered. Fig. 4a shows a snapshot of the application form for the registration process.

2- Once the users have completed the registration process, they can enter into the position tracking model. The tracking model is to send user's position information into the database of the system as well as shows the google map regarding to their positions, as shown in Fig. 4b.

3- Beside this, the users are also can receive the notification or alert about the areas which have been visited by infected users. The notification is working in the background, i.e. the user may be paused the application and uses other application on the smartphone. However, when the user opens the application and enters the infected area will receive the alert dialog. Fig. 4c and Fig. 4d show an example of the notification and alert dialogue. The notification and dialogue alert models are also configure both outdoors and indoors. For example, for outdoors, the GNSS position information of the users is used to measure the distance between any two users' positions and then if the distance is less than 5 meters then the notification or the alert dialog would be raised. However, for indoors, the application scans for Bluetooth devices in the vicinity, and then the result of the scan is matching with pre-registered MAC addressed in the system. If the matched MAC addresses have COVID-19 or recovered cases then the notification model and the alert dialog will notify the users about having COVID-19 or recovered users in the scan area.

### B. System Dashboard/Portal

A web portal for the system's administrators is designed and implemented using HTML5, PHP, JAVAScript, and google Map API. This part of the system is to monitoring and tracing the registered users only in terms of how the areas (which have been visited by users) should be lockdown or not? To this end, an unsupervised machine learning (UML) algorithm has been implemented in the system. There are several UML algorithms including Neural networks, Anomaly detection, Clustering and etc. However, for this system, K-means Clustering algorithm is used to predict the lockdown approach for the visited area. The K-means algorithm, first, reads the tracked users' position information and their status COVID-19. Then, in the next step will calculate the centroid position of the areas based on the DASV seeding method. The DASV method is a good algorithm to select the best centroid position of a set of nearest positions in the vicinity. Then, the centroid positions will be updated based on how the positions are nearest to each them. The pseudo code of the K-means clustering algorithm is shown in Algorithm 1.

---
**Algorithm 1** K-Means Clustering
---
**Input:** A specific region r & position information $x$
**Output:** cluster k
1 **while** $k_1, k_2, k_3...k_j \in r$ **do**
2    **for** $i \leftarrow x_1$ **to** $x_i$ **do**
3       -find the nearest centroid($c_1, c_2..c_k$)
4       -assign the position to that cluster
5    **for** $j \leftarrow k_1$ **to** $k_N$ **do**
6       new centroid = mean of all positions assigned to that cluster
---

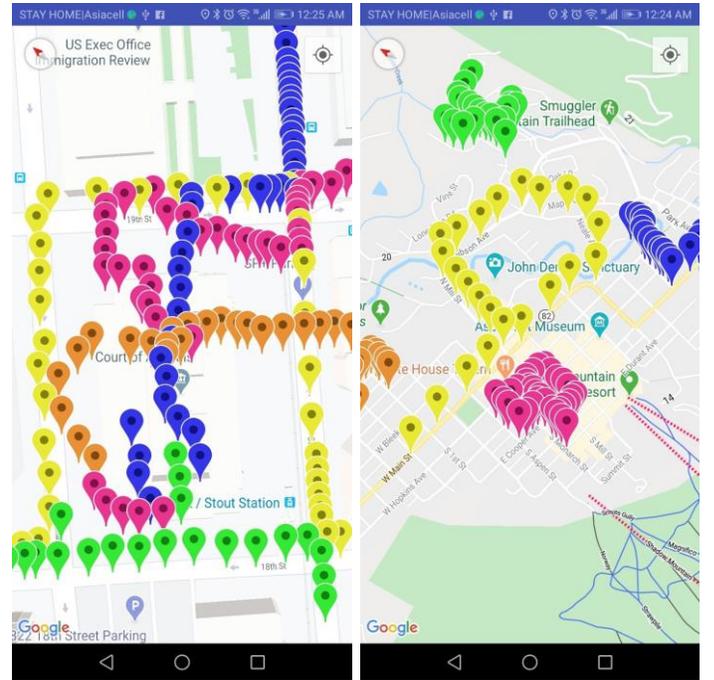

(a) Scenario 1      (b) Scenario 2

Fig. 6. Tracing five users in two different scenarios

Once, the process of the clustering of the tracked users' positions information has completed, a set of clusters will be produced. Then for each cluster, the distances between the positions of the different users are calculated. This is to calculate how many times the users, in the vicinity, are approaching to each other (from now called AEO). For this study, five users (user A, user B, user C, user D, and user E) are participated into the system in two different areas in USA. Therefore, two different scenarios via the five users are conducted for the K-means algorithm, as shown in figure 3. In the first scenario the users are walking and they are located in Denver area in Colorado-USA, while in the second scenario they are located in Aspen area in Colorado-USA.

A threshold for the approaching distance has been initialized to 5 meters, i.e. if user A has been approached around 5 meters to user B, or C, or D, or E, it means the users are too near to other users. For the two scenarios, if AEO is greater than 10, the system assumes this area is too crowded and the system will predict that the area should be locked down. However, if the value of AEO is less than 10 times, it means the area should not be locked down.

For 10 trial experiments, the model predicts that the Denver area in the first scenario should be locked down, since the five users during the walking in the area are approaching to each other for 55 times and they passed the threshold (i.e. 5 meters). However, in the second scenario, the same trials have been tested parallel with the second scenario, and the model predicted that the Aspen area doesn't need to be locked down, since the users are walked far to each other. Both scenarios results are shown in figure 4.

## V. Conclusion

At the emergence of COVID-19, many countries worldwide are commonly practiced social distancing, mass quarantine and even strict lockdown measures. Smart lockdown management is a pressing need to ease lockdown measures in places where people are practicing social distance. In this paper, we developed a smartphone-based approach to inform people when they are in proximity to an infected area with COVID-19. We also developed a dashboard to advise health authorities on how specific area safely get people back to their normal life. The proposed prediction model is used positional information and distance measurements of the registered users in the proximity. The government and public health authorities would be able to take benefit from the proposed dashboard to get latest statistics on COVID-19 cases and lockdown recommendation in different areas. The weak point of this study is the privacy issue of tracking position information of the users. This issue would be solved by applying encryption algorithms, in near future. However, such proposed system is significant to mitigate economic crisis and easing lockdown issues.

## VI. Compliance with Ethical Standards

Conflict of Interest: The authors declare that they have no conflict of interest. Moreover, this research was not funded by any funding agency.